%% file: main.tex
\documentclass[compsoc,conference,a4paper,10pt,times]{IEEEtran}
\IEEEoverridecommandlockouts
\usepackage{cite}
\usepackage{amsmath,amssymb,amsfonts}
\usepackage{algorithmic}
\usepackage{graphicx}
\usepackage{textcomp}
\usepackage{bmpsize}
\usepackage{xcolor}
\usepackage{lipsum}
\usepackage[colorlinks=true,urlcolor=black]{hyperref}
\def\BibTeX{{\rm B\kern-.05em{\sc i\kern-.025em b}\kern-.08em
    T\kern-.1667em\lower.7ex\hbox{E}\kern-.125emX}}

\usepackage{booktabs}
\usepackage{balance}
\usepackage{graphicx}
\usepackage{caption}
\usepackage{paralist}
\usepackage[inline]{enumitem}
\usepackage{xcolor}
\usepackage{multirow}
\usepackage{tabularx}   
\usepackage{textcomp}
\usepackage{subfig}
\usepackage{svg}

\usepackage{colortbl}
\usepackage{booktabs}
\usepackage{array}
\begin{document}

\title{From Beats to Breaches:\\How Offensive AI Infers Sensitive User Information from Playlists}

\author{
\IEEEauthorblockN{
Stefano Cecconello\IEEEauthorrefmark{1}, 
Mauro Conti\IEEEauthorrefmark{1}\IEEEauthorrefmark{3}, 
Luca Pajola\IEEEauthorrefmark{2}, 
Luca Pasa\IEEEauthorrefmark{1}, 
Pier Paolo Tricomi\IEEEauthorrefmark{1}}\\

\IEEEauthorblockA{\IEEEauthorrefmark{1}University of Padova, Italy\\
Email: \{stefano.cecconello, mauro.conti, luca.pasa, pierpaolo.tricomi\}@unipd.it}\\

\IEEEauthorblockA{\IEEEauthorrefmark{2}Spritz Matter, Italy\\
Email: luca.pajola@spritzmatter.it}\\

\IEEEauthorblockA{\IEEEauthorrefmark{3}Örebro University, Sweden}
}

\maketitle

\begin{abstract}
\input{sections/00-Abstract}
\end{abstract}


\input{sections/01-Introduction}
\input{sections/02-RelatedWork}
\input{sections/03-Background}
\input{sections/04-Framework}
\input{sections/05-Defense}
\input{sections/06-ExperimentalSettings}
\input{sections/07-Results}
\input{sections/08-Conclusions}
\balance
\section*{Acknowledgment}
This work was partially supported by the AISec4IoT project.
{

\bibliographystyle{IEEEtran} 
\bibliography{bibliography}   

\appendices

\end{document}

%% file: sections/00-Abstract.tex
The pervasive integration of \textit{Artificial Intelligence (AI)} has established a paradigm shift, giving rise to \textit{Offensive AI}, the exploitation of AI for malicious ends across the cyber-kill chain. A critical manifestation is the \textit{user attribute inference attack}, where AI infers sensitive \textit{Private Identifiable Information (PII)} from seemingly innocuous public data. We explore how modern music streaming ecosystems, where users routinely release public playlists without fully realizing the privacy implications, can be exploited for Offensive AI. To rigorously investigate and quantify this specific threat, we developed \textit{musicPIIrate}.
This novel Offensive AI tool leverages cutting-edge deep learning architectures and techniques that utilize not only standalone data representations but also the structural information embedded in a user's playlist collection and its interconnections. Our analysis explores a range of solutions, from set-based approaches such as pooling and Deep Sets (Deepset) to more advanced methodologies capable of modeling relationships between playlists, such as Graph Neural Networks (GNNs), which we also combined to leverage both perspectives.
Our design addresses the core challenge of feature extraction from unordered, variable-length set data, enabling highly efficient and accurate PII prediction.
\par
Our empirical evaluation demonstrates that musicPIIrate achieves \textit{state-of-the-art inference accuracy} in the domain of PII prediction from music playlist data. The tool successfully infers a wide array of user attributes, including: \textit{Demographics} (Age, Country, Gender, etc.), \textit{Habits} (Alcohol, Smoke, Sport), and \textit{Personality Traits} (OCEAN scores). Notably, musicPIIrate significantly outperforms existing methods, beating the baselines in 9 out of 15 attribute inference tasks.
To counter this significant privacy vulnerability, we further propose and implement \textit{JamShield}, a practical and lightweight \textit{defensive framework}. JamShield operates by strategically injecting \textit{dummy playlists} into a user's account to dilute the PII-carrying signal. 
Our analysis indicates that \textit{JamShield} represents a promising line of defense, lowering its F1-score by an average of 10\%.
This work provides an initial exploration toward an Offensive-AI benchmark for playlist-based PII inference using architectures that leverage both set- and graph-structured data (e.g., Deepset/GNN) and introduces a preliminary defense that shows encouraging mitigation effects.

%% file: sections/01-Introduction.tex
\section{Introduction}
During the past decade, Artificial Intelligence (AI) has rapidly transitioned from a research novelty to a pervasive technological cornerstone, enabling applications in healthcare, finance, transportation, cybersecurity, and critical infrastructure. As with any transformative technology, AI itself remains fundamentally neutral~\cite{schroer2025exploiting}: its impact is not defined by its underlying mechanisms, but by how it is used. Although much of the focus has been on harnessing AI for societal benefit, a parallel and significantly underexamined trajectory has emerged: the exploitation of AI for malicious ends. We refer to this adversarial paradigm as \textbf{Offensive AI}~\cite{schroer2025sok}. 
\par 
A notorious example of offensive AI is the \textbf{deep fake}, often utilized to spread misinformation and influence public opinion~\cite{farid2022creating}. Similarly, with the increased utilization of Large Language Models (LLMs), the problem of \textbf{spear phishing} is rising, as more individuals can easily customize massive phishing campaigns. A recent FBI report highlights a significant escalation in this threat, as criminals are now leveraging generative AI to produce more believable and scalable fraudulent schemes~\cite{IC3_2024}. This enables them to bypass common indicators of fraud and target a wider audience with enhanced deception.
\par 
The use of AI in cybercrime is not limited to a single phase of an attack. AI-based tools are now spreading across the entire \textbf{cyber-kill chain}, allowing adversaries to automate and accelerate their operations from reconnaissance to data exfiltration~\cite{schroer2025exploiting}. This means that even criminals with limited technical knowledge can now execute highly sophisticated, targeted attacks with speed and efficiency. Consider \textit{MalGan}~\cite{hu2022generating}: this tool is based on a generative model to produce malware that can bypass detector mechanisms. 
\par 
The growing prevalence of AI in cybercrime extends beyond the execution of the attack, encompassing a new frontier of reconnaissance. A \textbf{user attribute inference attack} is a privacy breach in which an adversary uses AI to infer user-sensitive personal information (e.g., gender, age)~\cite{adali2012predicting, gong2018attribute}.
Attackers can therefore infer from public information attributes that users intend to keep private, and be used, for instance, to launch spear phishing campaigns with frameworks like \textit{E-PhishGen}~\cite{pajola2025phishgen}.
Within this attack scenario, recently, researchers have proved the feasibility of inferring Private Identifiable Information (PII) from Spotify by leveraging public playlists available in users' accounts~\cite{tricomi2024all}.

\subsubsection{Contributions}
In this work, we expose a previously underestimated privacy risk in modern music streaming ecosystems by demonstrating that public playlists can be systematically weaponized through Offensive AI to infer sensitive user information. We develop an advanced deep learning pipeline capable of extracting PII-carrying behavioral signals from unordered and variable-length playlist sets and show that this threat is not merely theoretical, but highly practical and alarmingly accurate in real-world conditions. To prevent the exploitation of this vulnerability, we further design and evaluate a lightweight and deployable defensive strategy that effectively mitigates inference attacks without disrupting user experience.
We summarize our contributions as follows.
\begin{itemize}
    \item musicPIIrate, an Offensive AI framework for playlist-based PII inference. We explore architectures that leverage set- and relational representations of playlists and find that combining Deep Sets and Graph Neural Networks yields the best performance. This design models playlists as unordered sets while capturing complex track-level and playlist-level relationships, enabling high-precision PII prediction without manual feature engineering.
    \item A comprehensive empirical study establishing the feasibility and severity of the threat. musicPIIrate is evaluated across a broad range of user attributes—including demographics, personal habits, and personality traits—and achieves state-of-the-art results, outperforming baselines in 9/15 inference tasks.
    \item JamShield, a practical and lightweight defensive mechanism. We propose a defense based on injecting dummy playlists to dilute sensitive behavioral patterns. JamShield consistently degrades inference performance by an average of 10 F1-score points, demonstrating that the PII signal embedded in playlists can be neutralized without altering user behavior.
    \item A reproducible benchmark for future research on Offensive AI and privacy defenses. The codebase is available upon request, supporting reproducibility and further investigation of playlist-based PII inference and mitigation techniques.
\end{itemize}
\subsubsection{Ethical Statement}
The research presented in this paper follows the principles of responsible disclosure. While we provide a detailed analysis of privacy vulnerabilities in \textit{OffensiveAI}, we recognize the potential for misuse. To mitigate this risk and prevent the weaponization of our findings, we have adopted a restrictive release policy. 

In accordance with community best practices, we are not making the full implementation publicly available in an open repository. Instead, the code and experimental artifacts will be released upon request to verified researchers and practitioners for academic or defensive purposes. Requests must be accompanied by institutional verification. 

%% file: sections/02-RelatedWork.tex
\section{Related Work}
In this section, we review prior research relevant to our study. We covered three main areas: the psychological significance of musical behavior, the methods for inferring user characteristics from music consumption, and the broader literature on attribute inference from public digital traces.

\subsection{Music as a Psychological Signal}

A long tradition of research has established that musical preferences serve as meaningful psychological markers. Foundational studies showed that listeners’ personalities, particularly their Big Five profiles, systematically relate to the genres they consume. For example, Rentfrow and Gosling~\cite{rentfrow2003re} demonstrated that Openness is associated with preferences for complex genres such as jazz and classical, whereas Extraversion aligns with energetic mainstream styles like pop or rock. Later work refined this picture by examining how personality shapes not only genre choice but also the emotional and functional uses of music~\cite{chamorro2007personality, sloboda2001emotions}. Age and cultural background also play an important role: North and Hargreaves~\cite{north2008social} reported consistent generational differences in taste (e.g., younger listeners preferring contemporary high-arousal genres) as well as cross-cultural divergences in musical meaning and valuation~\cite{rentfrow2012role}. Collectively, these studies suggest that musical behavior embeds a wide range of psychological and demographic signals.

\subsection{Inferring User Characteristics from Music Behavior}

With the growth of streaming platforms, an increasing number of studies have investigated whether user attributes can be recovered algorithmically from music consumption data. Early computational approaches relied primarily on Last.fm logs. Liu et al.~\cite{liu2012inferring} showed that demographic variables such as age and gender can be predicted from listening histories enriched with temporal and audio-derived features. Similarly, Krismayer et al.~\cite{krismayer2019predicting} inferred age, gender, and nationality from listening patterns using a combination of collaborative signals and content descriptors. 

More recently, research has moved beyond demographic inference. Anderson et al.~\cite{anderson2021just} used private Spotify data, listening histories, user metadata, and app usage, to connect personality traits with detailed behavioral patterns across genres, artists, and sessions. In parallel, new large-scale studies have begun to rely on \emph{public} signals. Tricomi et al.~\cite{tricomi2024all} analyzed thousands of publicly shared Spotify playlists and demonstrated that playlist-level characteristics encode a wide set of personal attributes, including lifestyle habits and personality traits. Sust et al.~\cite{sust2023personality} expanded this direction using both audio and lyric-based features extracted from tracks consumed in naturalistic settings, showing that personality facets can be predicted with non-trivial accuracy. Finally, Sah et al.~\cite{sah2025perfairx} incorporated inferred personality vectors into LLM-based music recommenders trained using Last.fm data, revealing benefits for personalization but also systematic fairness disparities.

Together, these studies illustrate that music consumption, whether derived from listening logs, content features, or curated playlists, provides rich information that can be leveraged for attribute inference.

\subsection{Attribute Inference from Public Digital Traces}

Outside the music domain, a broad literature on Attribute Inference Attacks (AIA) has demonstrated that publicly available digital traces can reveal sensitive personal information. Kosinski et al.~\cite{kosinski2013private} famously showed that Facebook Likes alone allow accurate prediction of attributes such as sexual orientation, political leaning, and personality. Other work has linked social graph properties~\cite{golbeck2011predicting}, public ratings~\cite{weinsberg2012blurme}, and diverse forms of online behavior, including emoji usage and gaming telemetry~\cite{tricomi2023attribute}, to private individual traits. These findings underscore the privacy risks posed by user-generated content, even when such content is not explicitly intended to convey personal information.

Research on music data fits naturally within this broader landscape: Liu et al.~\cite{liu2012inferring} and Krismayer et al.~\cite{krismayer2019predicting} already highlighted the sensitivity of listening histories, while more recent works such as Tricomi et al.~\cite{tricomi2024all} stress that even publicly accessible playlists may expose personal attributes. From a privacy perspective, attribute inference from music consumption can thus be viewed as a domain-specific instantiation of a more general class of inference attacks, where publicly visible behavior serves as an indirect proxy for sensitive traits.



%% file: sections/03-Background.tex
\section{Background}
This section introduces the concepts and methods used to formalize the proposed offensive framework. Throughout the text we adopt the convention that lowercase letters (e.g., $s$) denote scalars, bold lowercase letters (e.g., $\mathbf{x}$) denote vectors, and bold uppercase letters (e.g., $\mathbf{B}$) denote matrices.

\subsection{Graph Neural Networks}
\label{sec:GNNs}

Let $G=(V,E,\mathbf{X})$ be a graph where $V=\{v_1,\ldots,v_{n}\}$ is the vertex set, $E\subseteq V\times V$ is the set of edges, and $\mathbf{x}_{v_i}$ denotes the feature vector associated with node $v_i$. We write $\mathbf{A}\in\mathbb{R}^{n\times n}$ for the adjacency matrix of the graph, with entries $a_{ij}=1 \iff (v_i,v_j)\in E$. The neighborhood of a node $v$ is indicated by $\mathcal{N}(v)$.

A Graph Neural Network (GNN) is a model designed to leverage both the structural properties of a graph and the feature information associated with its nodes. Its goal is to learn an embedding (or representation) $\mathbf{h}_v \in \mathbb{R}^m$ for each vertex $v \in V$. Contemporary GNN architectures typically separate this computation into an \emph{aggregation} step and a \emph{combination} step. Using two functions $\mathcal{A}$ and $\mathcal{C}$ these operations can be written as
\begin{equation}
 \mathbf{h}_v=\mathcal{C}\bigl(\mathbf{x}_v,\; \mathcal{A}(\{\mathbf{x}_u: u\in\mathcal{N}(v)\})\bigr). \nonumber
\end{equation}

By repeating aggregation and combination for $k$ iterations one extends the receptive field to include nodes up to distance $k$. Denoting by $\mathbf{h}_v^{(i)}$ the node representation after $i$ iterations, we have:
\begin{align}
    \mathbf{h}_v^{(i)}&=\mathcal{C}\Bigl(\mathbf{h}_v^{(i-1)},\; \mathcal{A}(\{\mathbf{h}_u^{(i-1)}: u\in\mathcal{N}(v)\})\Bigr), \quad i\in[1,\dots,k],\nonumber\\ 
    \mathbf{h}_v^{(0)}&=\mathbf{x}_v,\qquad \mathbf{h}_v^{(i)}\in\mathbb{R}^{m_{i}}, \nonumber
\end{align}
where $m_i$ denotes the dimension of the convolutional layer $i$. In this way, we obtain a deep GNN of $k$-layers. The particular choices of $\mathcal{A}$ and $\mathcal{C}$ determine the Graph Convolution (GC) operator used by the network~\cite{micheli2009neural,kipf2016semi}.

The particular choice of the aggregation operator $\mathcal{A}$ together with the combination function $\mathcal{C}$ determines which kind of \emph{Graph Convolution (GC)} the GNN implements. Numerous GC variants have been introduced in recent years. A common design for $\mathcal{A}$ is to sum the embeddings of neighboring nodes. Since summation is commutative, this makes the aggregation insensitive to the ordering of neighbors. A widely used variant of GC oprtaor was proposed by Kipf and Welling~\cite{kipf2016semi}. Their Graph Convolutional Network (GCN) is expressed as:
\begin{equation}
    \mathbf{h}_v^{(i)}= \phi\Bigl(\mathbf{W}_{\Sigma}\sum_{u \in \{\mathcal{N}_v \cup v\}} \mathbf{h}_u^{(i-1)} +\mathbf{b}\Bigr). \nonumber
\end{equation}
Here $\mathbf{W}_{\Sigma}\in\mathbb{R}^{m_{i}\times m_{i-1}}$ (with $m_0 = s$) and $\mathbf{b}\in\mathbb{R}^{m_{i}}$ are learnable parameters.
A common design for the aggregation $\mathcal{A}$ is to sum neighbor embeddings. This commutative operation ensures invariance to the ordering of neighbors. 
Although the framework described above has achieved strong results in many different application domains, it also presents some important drawbacks.A major limitation of GNNs is the oversmoothing effect: as aggregation and update operations are repeatedly applied, node embeddings gradually converge to similar values. This diminishes the model’s ability to distinguish between nodes, reduces expressiveness, and ultimately degrades performance~\cite{Keriven_2022}. This issue becomes even more pronounced in deep architectures, where multiple stacked layers cause excessive information mixing and a loss of discriminative power.
While in recent years, especially in the context of structured data, there has been a growing trend toward adopting increasingly complex and deeper models to obtain richer, multi-level abstract representations of the input, the multi-resolution architecture\cite{pasa2022} also led to a simplification of the graph convolution operator, resulting in the Simple Graph Convolution (SGC)\cite{Wu2019}. SGC is built on the observation that the nonlinear transformations used in traditional GCNs may not be essential for effective learning. Instead, the authors suggest that stacking several linear graph convolution operators is sufficient to capture meaningful structure. The great advantage of this model is a reduced number of parameters compared to classical graph convolution.

\subsubsection{Pooling}
When dealing with graph-level prediction tasks, the node-level representations enriched by topological information must be combined to produce a single, fixed-size embedding for the entire graph. This aggregation step is fundamental, as it must convert a variable number of node representations into one graph-level vector. Furthermore, a good graph-level representation should ideally be invariant to any isomorphic relabeling of the input graph, ensuring that the learning process focuses solely on the target property without being influenced by the particular ordering or encoding of the nodes. This process of aggregating node-level embeddings into a graph-level representation is commonly referred to as \textbf{pooling}.
A widely used pooling strategy in many GNN architectures is to apply simple aggregation operators such as the mean, the element-wise maximum, or the sum. However, recent studies~\cite{Zaheer2017,Navarin2019} indicate that these basic aggregation methods can cause information loss because they merge numerical values in a manner that may obscure relevant distinctions, ultimately degrading the predictive performance of the model.

\subsubsection{Deepset}
To address the limitations of simple pooling strategies, one can instead treat the topologically enriched node representations as a set and learn an aggregation function that operates directly on sets. DeepSets~\cite{Zaheer2017} offer a principled framework for building neural architectures whose inputs are sets. In contrast to basic aggregation rules, the DeepSets formulation is maximally expressive: under appropriate assumptions, it can be shown to serve as a universal approximator for functions defined on sets. In this approach, each element of the input set is first mapped into a high-dimensional space through a learned function $\phi(\cdot)$—typically implemented as a multilayer perceptron~\cite{Navarin2019}. The resulting embeddings are then summed to produce a single vector that summarizes the set, and a readout function $\rho(\cdot)$ (also an MLP) transforms this summary into the final output for the task.
One of the main arguments of the universal approximation proof of DeepSets for the countable case, i.e., where the elements of the sets are countable ($|\mathfrak{X}| \leq n_0$), relies on the fact that, 
given the space of input sets $\mathcal{X}\subseteq 2^{\mathfrak{X}}$, any function
over sets can be decomposed as $sf(X)=\rho(e(X))$, where $e: \mathcal{X} \rightarrow \mathbb{R}^n$, $e(X)=\sum_{x_i \in X}\phi(x_i)$,  combining the elements $x_i\in X$ non-linearly transformed by the $\phi(\cdot)$ function, maps different sets in different points.
%
To ensure that $\rho(\cdot)$ can associate different outputs with different inputs, the mapping $\phi(\cdot)$ must encode individual set elements in such a way that $e(\cdot)$ yields a unique representation for each $X\in\mathcal{X}$. In the countable case, this can be achieved by choosing $\phi(\cdot)$ so that every element is mapped to a representation orthogonal to those of all other elements. In contrast, for uncountable domains the construction is more involved and requires $\phi(\cdot)$ to satisfy a homomorphism property.
Building on this theoretical foundation, Navarin et al.~\cite{Navarin2019} introduce a graph-level aggregation mechanism inspired by DeepSets, implementing the function $\phi(\cdot)$ as a multilayer perceptron.

%% file: sections/04-Framework.tex
\section{Offensive Framework: musicPIIrate}\label{sec:musicPIIrate}
We introduce \textit{musicPIIrate}, our framework for performing attribute inference attacks on publicly shared music playlists. In the following subsections, we formalize the threat model, define the structure of a music-streaming platform, and describe the machine learning classifiers used to infer user attributes from playlist data.

\subsection{Proposed Threat Model}
\label{ssec:definition}

Our Attribute Inference Attack (AIA) targets users of online music-streaming platforms (e.g., Spotify, Apple Music) who publicly share their playlist collections. Modern platforms encourage public playlist creation and sharing across social media, making curated music content widely accessible. For simplicity, we assume that each user has a unique platform identifier (e.g., profile link or username) that can be used to retrieve publicly available playlists and their metadata.

\textbf{Formal Definition.}
We characterize the attacker's perspective by following four standard criteria:
\begin{itemize}
    \item \textit{Goal:} The attacker aims to infer personal attributes of a target set of users (e.g., age group, gender, sexual orientation, political preference, or mental health indicators), even though these attributes are never disclosed on the platform.
    
    \item \textit{Knowledge:} The attacker knows the public identifier (e.g., username or profile URL) of each target user and is familiar with the music-streaming ecosystem. The attacker is aware of the correlations between music consumption patterns and demographic or psychographic traits.
    
    \item \textit{Capability:} The attacker can access only public and voluntarily shared information. This includes playlist titles, track lists, song metadata (e.g., artists, genres, release dates), and platform-derived engagement statistics (e.g., likes or followers of a playlist). The attacker does \textbf{not} perform data breaches, exploit vulnerabilities, or access non-public datasets.
    
    \item \textit{Strategy:} First, the attacker legitimately collects a training dataset comprising public playlists from users whose personal attributes can be unobtrusively obtained through secondary public sources (e.g., social networks, public fan communities, online surveys). The attacker then trains a machine learning model to map playlist-derived features to personal attributes. Finally, the attacker applies this model to infer the private attributes of users whose personal information is not explicitly disclosed, relying solely on their public playlists.
\end{itemize}

We emphasize that users voluntarily publish their playlists for social interaction or self-expression and are unlikely to realize that such information may enable the inference of highly sensitive personal traits. The attack requires neither unauthorized access nor incentive to perform a breach, as inference at scale is already feasible using publicly available data.

\begin{figure*}[h]
  \centering
  \includegraphics[width=\linewidth]{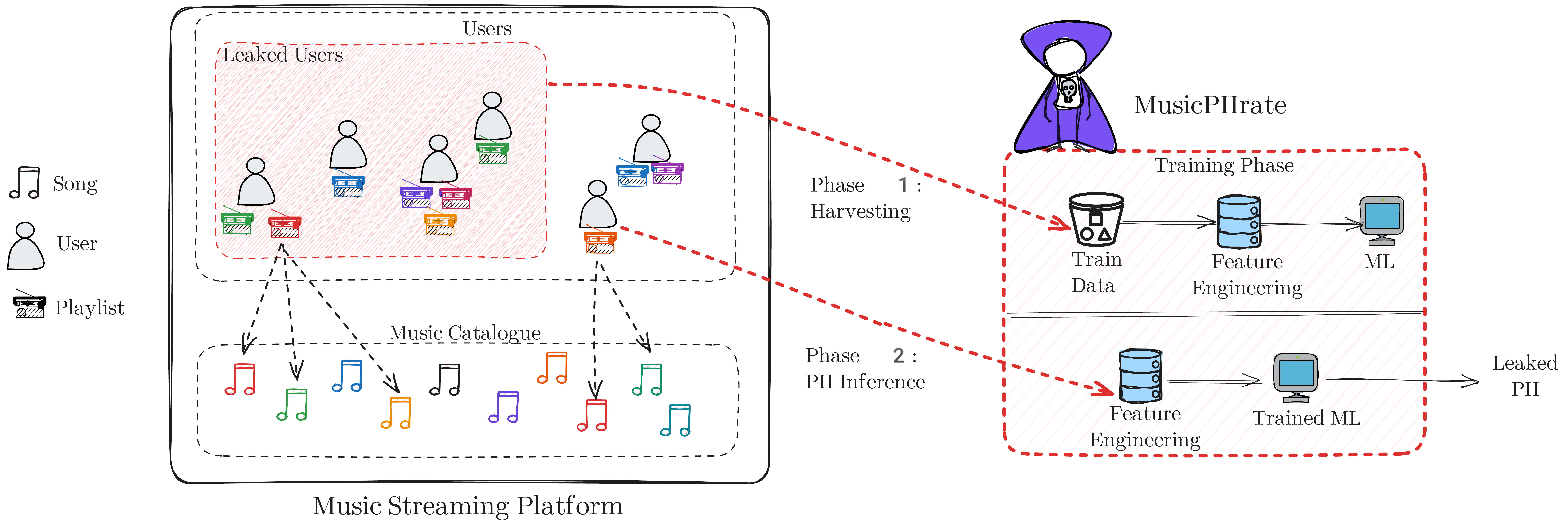}
  \caption{\textit{Architecture of the MusicPIIrate Privacy Attack.} The framework harvests public music playlist data from streaming platforms (Phase 1) to train an ML model. This model then exploits feature-engineered music preferences (Phase 2: PII Inference) to extract and leak Personally Identifiable Information (PII) from other users.}
\end{figure*}

\subsection{Mathematical Modeling of a Music-Streaming Platform}
\label{ssec:platform_model}

We now formalize the structure of a music-streaming platform to support the analysis of our AIA. Let \mbox{$\mathcal{U} = \{u_1,\dots, u_{n}\}$} denote the set of platform users and $\mathcal{S}$ the global catalog of songs available on the platform. Each user $u \in \mathcal{U}$ publicly maintains a collection of playlists. We denote the set of playlists of user $u$ as
\[
\mathcal{P}_u = \{P_1^{(u)}, P_2^{(u)}, \dots, P_{n_u}^{(u)}\},
\]
where $n_u$ is the number of playlists published by the user $u$. Each playlist $P_i^{(u)}$ is an ordered list of songs, i.e.,
\[
P_i^{(u)} = \big(s_{i,1}^{(u)},\, s_{i,2}^{(u)},\, \dots,\, s_{i,k_i}^{(u)} \big),
\]
with $s_{i,j}^{(u)} \in \mathcal{S}$, $j \in [1, \dots, k_i]$ and $k_i$ the number of songs contained in playlist $P_i^{(u)}$.
Every song $s \in \mathcal{S}$ is associated with a structured metadata tuple:
\[
\xi(s) = \big( \text{artist}(s),\ \text{genre}(s),\ \text{release\_year}(s),\ \text{mood}(s),\ \dots \big).
\]
We assume that this metadata is fully public, either directly on the platform or through third‑party catalogs exposed by the service.

For each user $u$, we define the \emph{listening signature} as the multiset of metadata aggregated from their public playlists:
\[
\Xi^{(u)} = \biguplus_{P_i^{(u)} \in \mathcal{P}^{(u)}} \biguplus_{s \in P_i^{(u)}} \xi(s),
\]
where $\biguplus$ denotes multiset union, preserving repeated occurrences. $\Xi^{(u)}$ provides a high‑dimensional representation of the user's musical preferences in terms of artists, genres, moods, and temporal trends.

Finally, we let $\mathcal{T}^{(u)}$ represent the (unknown) 
private personal attributes of user $u$, such as age group or gender. For a subset of users $\mathcal{U}_{\text{train}} \subset \mathcal{U}$, the attacker is able to obtain $(\Xi^{(u_i)}, \mathcal{T}^{(u_i)} \;| \; \forall u_i \in  \mathcal{U}_{\text{train}})$ pairs from publicly available secondary sources (e.g., self-declared information on social networks). For the remaining users $\mathcal{U}_{\text{target}} = \mathcal{U} \setminus \mathcal{U}_{\text{train}}$, the personal attributes $\mathcal{T}^{(u_j)}, \; u_j \in \mathcal{U}_{\text{target}} $ are unknown, while their playlists and thus $\Xi^{(u_j)}$ are public.

In summary, the proposed framework, given a user $u$ and their listening signature $\Xi^{(u)}$, aims to model a function able to derive their sensitive attributes $\mathcal{T}^{(u)}$.

\subsection{Attribute Inference Classifiers}


The proposed approach relies on an ML classifier to infer a user’s sensitive attribute from the user’s playlists. In our framework, we evaluate three different kinds of approaches, which differ in the structure of the input data. 
The goal of the models we define is to classify a single user. Each user is represented by a set of playlists denoted as $\mathcal{P}^{(u)}$. Each playlist in this set is represented by a vector $\mathbf{p}_i^{(u)} \in \mathbb{R}^m$.
Here we consider three different approaches based on how the model handles the relationships among the playlists $\mathcal{P}^{(u)}$. In the first approach, we define a model that learns to map a single playlist to the correct class. A user is then classified by applying the same model multiple times, once for each of the user’s playlists, and averaging the resulting predictions. We refer to these models as the \textbf{sample-based approaches}. This technique is similar to the method proposed in \cite{tricomi2024all}, which we use as a baseline.
The second class of models we developed learns how to classify the entire set $\mathcal{P}^{(u)}$. These models, which we refer to as \textbf{set-based approaches}, exploit pooling methods and the DeepSets paradigm to compute a set level representation. More precisely, the model uses an MLP that projects each playlist $\mathbf{p}_i^{(u)}$ into a latent space, and then aggregates the resulting representations by means of a pooling technique or a DeepSet architecture.These models, unlike the sample based approaches, are trained in an end-to-end fashion.
Finally, the last type of ML techniques we developed are the \textbf{graph-based approaches}. Unlike the previous methods, these models take as input a graph whose nodes correspond to playlists and whose edges are computed according to a similarity measure between playlists. The graph contains all playlists in $\mathcal{P} = \bigcup_{u \in \mathcal{U}} \mathcal{P}^{(u)}$. This class of approaches applies graph convolutions on the graph, computing for each node a richer embedding that captures information from its neighboring playlists. The nodes associated with each user are then aggregated using a subset pooling technique that combines all nodes belonging to that user. As an alternative to the pooling technique, we also propose and developed a DeepSubSet approach, which applies the same principle as DeepSets but is restricted to the subset of graph nodes belonging to a specific user.

Exploring these three types of approaches is motivated by the fact that they operate at different informational levels. The sample-based approaches are essentially local learning methods that have access only to a single playlist at a time. Consequently, classification relies solely on local information and does not exploit the complete listening signature of the user. This limitation is addressed by set-based approaches, which consider all playlists of a user simultaneously, enabling the model to capture a more holistic and user-level representation. However, set-based methods still isolate each user, ignoring the fact that playlists often share songs across different users, revealing cross-user relationships. Graph-based approaches overcome this limitation by representing all playlists from all users as nodes in a graph, where edges connect playlists based on a similarity metric. This structure allows the model to compute embeddings that capture not only the user's own listening behavior but also the relationships and similarities with other users' listening signatures.
\subsubsection{Sample-based approaches}
This kind of approaches consider simple classifiers to address the task of attribute inference. The simple classifier $\mathcal{C}() $ process single playlist at a time, to produce a classification (probability) $y_i^{(u)}$ for each of them. We then combine the outputs of all the user's playlists and calculate the average classifications $y^{(u)}$ for the user:
\begin{equation}
    y^{(u)}=\frac{\sum_{\mathbf{p}_i^{(u)} \in \mathcal{P}^{(u)}}  \mathcal{C}(\mathbf{p}_i^{(u)})} {|\mathcal{P}^{(u)}|}. \nonumber
\end{equation}

\subsubsection{Set-based approaches}
In set-based approaches, the model takes as input the set of playlists of user $u$, $\mathcal{P}^{(u)}$, and learns to map this set directly to the classification $y^{(u)}$. This is achieved by first applying an MLP, $\mathcal{M}()$, to each element in $\mathcal{P}^{(u)}$ and then applying an aggregation function, $\mathcal{A}()$, over the resulting representations:
\begin{equation}
y^{(u)} = \mathcal{A}(\{ \mathcal{M}(\mathbf{p}_i^{(u)}) | \; \forall \; \; \mathbf{p}_i^{(u)} \in \mathcal{P}^{(u)} \}), \nonumber
\end{equation}
where $\mathcal{A}()$ can be a pooling operator or a DeepSet. We refer to these two approaches as \textbf{MLP-pooling} and \textbf{MLP-deepset}, respectively.

\subsubsection{Graph-based approaches}\label{sec:graph_definition}
The class of methods we consider is the graph-based ones.
The input of the model is a graph  $G_p = (\mathcal{P}, \Upsilon ,\mathbf{P})$ where the nodes correspond to all playlists of all users. The node attribute matrix is
$$
\mathbf{P} = \begin{bmatrix} \mathbf{p}_0^{(u_1)}\\
\mathbf{p}_1^{(u_1)}\\
\vdots\\
\mathbf{p}_i^{(u_j)}\\
\vdots\\
\end{bmatrix},$$
where $i \in [1,\dots,n_u],\; j \in [1, \dots n],$ and, $ \mathbf{p}_i^{(u_j)}$ denotes the feature vector of the $i$-th playlist of user $u_j$. The edge set is defined as: 
$$
\Upsilon=\{e_{\mathbf{p}_i, \mathbf{p}_j}|\mathbf{p}_i, \mathbf{p}_j\in \mathbf{P} ,\; \exists e_{\mathbf{p}_i,  \mathbf{p}_j} \iff \Psi(\mathbf{p}_i, \mathbf{p}_j) \ge \tau \}
$$
that is, an edge exists between two playlists $\mathbf{p}_i$ and $\mathbf{p}_j$ if the measure $\Psi(\mathbf{p}_i, \mathbf{p}_j)$ is greater than or equal to a predefined threshold $\tau$.
The models considered here share the same structure, which consists of a multilayer GNN (we refer to this stack of GCN layers as $\mathcal{G}$) that provides richer embeddings for each node of the graph (without modifying the graph topology). The resulting node embeddings are treated as a set and passed to a user-aware aggregation layer $\mathcal{A}()$, which aggregates the nodes belonging to each user separately and produces the average classifications $y^{(u)}$ for each user $u$ as a column vector $\mathbf{y}=\begin{bmatrix}y^{(u_1)},\\ \vdots \\ y^{(u_{n_u})}  \end{bmatrix}$.
Formally, the resulting model, hereafter referred to as \textbf{GNN-pooling}, is defined as follows:
\begin{equation}
    \mathbf{y}=\mathcal{A}(\{\mathcal{G}(G_p)\}, \mathcal{U}, \mathcal{P}).\nonumber
\end{equation}
The second methodology we introduce is the DeepSubSet, which is an extension of DeepSet able to act on a subset of the nodes of the graph. To achieve this, we modified the inner aggregation function of the original DeepSet, which is applied on the output of $\phi$.
The novel inner aggregation exploits not only the node embeddings but also the information necessary to identify the subset of belonging for each graph node ($\mathcal{U}$ and $\mathcal{P}$). In this way, $\phi$ computes a transformation of the input embeddings, and subsequently, the inner aggregation function takes care of passing the $\rho$ function a vector of the sub-set level embeddings
\begin{equation}
    \mathcal{A}(G, \mathcal{U}, \mathcal{P}) =\rho\!\left( \left[\sum_{p_i^{(u_j)}\in P^{(u_j)}}\phi(p_i^{(u_j)})\right]_{u_j \in \mathcal{U}}  \right),
    \label{eq:deepset}
\end{equation}
We refer to this model as \textbf{GNN-DeepSet}.

%% file: sections/05-Defense.tex
\section{Defensive Framework: JamShield}
In this section, we introduce \emph{JamShield}, a defensive framework designed to mitigate the Attribute Inference Attack (AIA) presented in the previous section and to improve the privacy of users of music-streaming platforms against inference from publicly shared playlists. We propose investigated three countermeasure that operates at two different levels: (i) \emph{feature-level} countermeasures, which directly modify or remove playlist features used by an attacker, and (ii) \emph{data-level adversarial augmentation}, which intentionally augments a user's publicly visible playlist collection with carefully chosen playlists whose purpose is to confuse and degrade the performance of the attacker's inference model. Below, we present the conceptual design of JamShield.

\subsection{Feature-level countermeasures: noise and ablation}
The first family of JamShield defences works by intervening on the features that compose the playlist embeddings used by the attacker. Recall that each user $u$ is represented by a set of playlists $\mathcal{P}(u)$, where $\mathbf{p}_i^{(u)}\in\mathbb{R}^m$ denotes the feature vector (embedding) of the $i$-th playlist. Feature-level defences do not alter the topology of playlists or the set membership, but act on the components of these vectors.

The first approach is \emph{additive noise}: each modifiable playlist embedding is replaced with a perturbed version
\[
\hat{\mathbf{p}}_i^{(u)} \;=\; \mathbf{p}_i^{(u)} + \boldsymbol{\delta}_i^{(u)}, \qquad \text{with}\;\; \|\boldsymbol{\delta}_i^{(u)}\| \le \varepsilon,
\]
where the perturbation $\boldsymbol{\delta}_i^{(u)}$ is chosen to increase the attacker’s uncertainty while keeping the modified embedding within a $\varepsilon$ so that the resulting embedding remains plausible and realistic.
The second feature-level technique is \emph{feature ablation}. Under ablation, a subset of features \(\xi^{(u)}\) is removed from the publicly released embeddings for user \(u\). Formally, let \(\mathbf{B}_{\xi^{(u)}}\) be the diagonal mask that zeroes out the coordinates corresponding to \(\xi^{(u)}\); the ablated playlist embedding is
\[
\hat{\mathbf{p}}_i^{(u)} \;=\; \mathbf{B}_{\xi^{(u)}} \, \mathbf{p}_i^{(u)}.
\]
Whereas additive noise perturbs the full embedding while preserving all coordinates, ablation eliminates the information contained in the selected dimensions entirely.

Note that for both feature-level defences, not all features can be safely modified or removed. Indeed, it is useless for the defense to modify or remove those features that an attacker can recover from external sources, like the publication year or the duration. See Section~\ref{sec:jamshield_exp} for the selection of the modifiable feature set. Both strategies are conceptually simple, inexpensive to deploy, and can be applied locally by the platform at export time. Their main limitation is that attackers can often adapt to simple noise distributions. These limitations motivate us to investigate also the stronger data-level countermeasure described next.

\subsection{Data-level countermeasures: playlist injection}
The central pillar of JamShield is a data-level defence that we call \emph{adversarial playlist injection} (hereafter simply \emph{injection}). Instead of only perturbing numeric features, injection operates directly on the publicly visible \emph{set} of playlists by adding one or more \emph{decoy playlists} whose purpose is to reduce the success of an attacker using the classifiers described in Section~\ref{sec:musicPIIrate}. Denote by \(\mathcal{J}(u)\) the set of injected playlists for user \(u\). The collection of playlists released publicly becomes
\[
\widehat{\mathcal{P}}(u) \;=\; \mathcal{P}(u) \;\cup\; \mathcal{J}(u).
\]
Because our target models reason over playlist collections (sample-, set-, and graph-based aggregators), injection changes the distribution of items in the input set rather than directly modifying individual embedding coordinates.
In practice, injection can be implemented by (i) creating and publishing new playlists under the user's account, (ii) promoting existing but less revealing playlists so they become more prominent, or (iii) leveraging platform-supported sharing constructs (e.g., collaborative playlists) to surface playlists that dilute the informative signals present in \(\mathcal{P}(u)\).
In practice, injection should be implemented at the platform level. When an external party queries the service’s API for a user’s public playlists, the platform should return the actual playlists \(\mathcal{P}(u)\) together with the decoy set \(\mathcal{J}(u)\), without revealing which playlists are genuine and which are injected. This ensures that attackers cannot distinguish between authentic and synthetic elements of the released set. The platform may additionally expose this mechanism as an opt-in privacy feature, allowing users to choose whether they prefer to publish their playlists as-is or enable JamShield’s decoy injection to increase protection against attribute inference.

We perform the injection by directly targeting the attacker’s loss: given the attacker model \(f^\star\) and the listening signature \(\Phi(\cdot)\), JamShield selects injected playlists \(\mathcal{J}(u)\) to maximize the loss \(L\big(f^\star(\Phi(\mathcal{P}(u)\cup\mathcal{J})),T(u)\big)\). The concrete optimization procedure, implementation details (including the exact surrogate used for \(f^\star\)), and the results of applying this loss-maximization strategy are reported in the experimental section (Section~\ref{sec:jamshield_exp}).


%% file: sections/06-ExperimentalSettings.tex
\section{Experimental Settings}\label{sec:ExperimentalSettings}
This section presents the experimental setup used to evaluate the musicPIIrate attack framework and the JamShield defensive mechanisms. 
We describe the dataset and target attributes, detail the baseline, set-based, and graph-based models, and outline the training procedures and evaluation metrics. 
Finally, we present the concrete implementation of both feature-level and data-level defenses in JamShield, ensuring a reproducible experimental framework.

\subsection{Dataset}\label{sec:dataset}
Our experiments are conducted on the dataset collected by Tricomi et al.~\cite{tricomi2024all}, which provides a rich collection of users' playlists and associated attributes.\footnote{All attributes are discrete. In our experiment, we refer to only the classification task. }
The dataset contains over 10,000 playlists shared by 739 users, spanning over 200,000 songs and 55,000 artists. The average number of playlists per user is 13.92.
In our notation, each user $u$ is represented by a set of playlists $\mathcal{P}(u)$, and each playlist $ \mathbf{p}_i^{(u)} \in \mathbb{R}^{111}$ is represented by a feature embedding. These embeddings form the basis for both attribute inference attacks and our defensive framework.

Each playlist embedding $\mathbf{p}_i^{(u)}$ is constructed from 111 features originally computed in the dataset. For each song in the playlist, summary statistics such as \textit{average} (avg), \textit{minimum} (min), \textit{maximum} (max), \textit{standard deviation} (std), and \textit{ratio} are computed for features where applicable. These statistics are aggregated across all songs in the playlist to form the embedding vector, capturing a wide range of musical, popularity, and structural characteristics.

In addition to the original playlist embeddings provided by the dataset, we define a playlist graph $\mathcal{G}_p$, following the graph-based approaches described in 
Section~\ref{sec:graph_definition}. Edges between playlists are defined by the function $\psi$, which in our experiments is instantiated as the intersection, returning the number of songs common to the two playlists. Furthermore, we set $\tau = 0$, so an edge is created between any two playlists that share at least one song.

The dataset provides a total of 16 target classes, divided into three categories. \textbf{Demographics} containing Age, Country, Economy, Gender, Marital Status, Occupation, and Live with. \textbf{Habits} containing Alcohol, Smoke, Sport, and Premium Account. \textbf{Personality Traits} containing the OCEAN scores, including Openness, Conscientiousness, Extraversion, Agreeableness, and Neuroticism. Analyzing the dataset, we observed that the \textit{Live with} class was extremely imbalanced, with one of the two categories having a negligible number of samples. Consequently, we excluded this class from our experiments, as no meaningful learning could be achieved even when applying oversampling or undersampling strategies.

\subsection{Baselines: Sample-based Approaches}
For comparison with our proposed methods, we evaluate a set of sample-based baseline models that share the same overall approach but differ in the specific classifier $\mathcal{C}$ employed. In particular, we consider logistic regression, decision trees, random forests, k-nearest neighbors (KNN), and multi-layer perceptrons (MLP). The hyperparameters for each classifier are selected through an internal validation procedure, as detailed in Section~\ref{sec:SelectionEvaluation}.
To address the class imbalance present in the dataset, oversampling is applied to the training data, augmenting underrepresented classes and improving each classifier’s ability to learn balanced decision boundaries while preserving the relative structure of the dataset. This combination of per-playlist modeling, user-level aggregation, and class balancing provides a robust baseline for evaluating the effectiveness of our privacy-preserving interventions.

\subsection{MusicPIIrate implementation}\label{sec:musicPIIrateImplementation}


In this section, we provide a detailed description of the concrete implementation of the musicPIIrate framework. We describe how each component of the pipeline is implemented in practice, specifying the encoders, aggregation operators, and training procedures used in both the set-based and graph-based variants. All architectural hyperparameters are validated as in Section~\ref{sec:SelectionEvaluation}.

\subsubsection{Set-Based Approaches}
Set-based models process all playlists of a user jointly to produce a single prediction. Each playlist embedding $\mathbf{p}_i^{(u)}$ is first projected into a latent space via a shared MLP encoder $\mathcal{M}$ that consists of multiple linear layers with ReLU activations, batch normalization, and dropout.

User-level predictions are then obtained using two alternative aggregation operators $\mathcal{A}$:
\begin{itemize}
    \item \textbf{MLP–Pooling:} The aggrgation function $\mathcal{A}$ is defined as concatenation of three festure leval function: sum, mean and max.
    \item \textbf{MLP–DeepSet:} In the DeepSet both $\phi$ and $\rho$ are implemented as separate MLPs with ReLU activations and dropout. We also applied to the representations a batch normalization. 


\end{itemize}



\subsubsection{Graph-Based Approaches}
Graph-based models operate on the graph $G_p = (\mathcal{P}, \Upsilon, \mathbf{P})$ introduced in Section~\ref{sec:dataset}, where each node corresponds to a playlist. The encoder $\mathcal{G}$ is instantiated as an SGC model, which computes enriched node embeddings that are subsequently aggregated using $\mathcal{A}$ to produce user-level predictions.

Two aggregation strategies are considered, analogous to those used in the set-based approaches:
\begin{itemize}
    \item \textbf{Standard pooling:} A user-level pooling operation that concatenates the outputs of three feature-wise aggregation functions: sum, mean, and max.
    \item \textbf{DeepSubSet:} this method reuses the same $\phi$ and $\rho$ MLP architecture as in the set-based variants. The key difference lies in the aggregation: node embeddings belonging to the same user subset are summed before being passed to $\rho$, as formalized in Eq.~\eqref{eq:deepset}.
\end{itemize}

\subsubsection{Training Protocol}
All models set-based and graph-based are trained with the same procedure: class-weighted negative log-likelihood loss, Adam optimizer, and hyperparameters (MLP/SGC depth, hidden units, dropout, learning rate, and architectures of $\phi$ and $\rho$) tuned via internal validation as described in Section~\ref{sec:SelectionEvaluation}.

Unlike the sample-based baseline, which classifies each playlist independently using a single shared class label and aggregates predictions post-hoc, both set- and graph-based models optimize directly at the user level. This ensures that the encoders $\mathcal{M}$ or $\mathcal{G}$ learn representations tailored for whole-user attribute inference.

\subsection{Experimental Validation}\label{sec:SelectionEvaluation}



For all models, both classical machine learning baselines and deep learning candidates, we enforce a 5-fold external cross-validation strategy to provide robust performance estimates. Splits are performed at the user level, ensuring that all playlists of a given user are contained entirely within a single fold. This design reflects realistic deployment scenarios, where attributes of users in the training set should not leak into the validation or test sets. Furthermore, splits are stratified according to the target attribute to maintain consistent class distributions across training, validation, and test sets.

For deep learning models, each fold is trained with 10 independent repetitions to mitigate the impact of variance and stochasticity in training. Model selection for these models is performed by monitoring the mean validation loss across the ten repetitions within each fold. Reported test performance for each fold is computed as the mean test F1 score over these ten repetitions.
Similarly, a grid search for baseline models is performed with 10 repetitions to allow fair comparison.

We adopt the macro F1 as the primary evaluation metric for deep learning models, as it assigns equal importance to all classes irrespective of their frequency, providing a fairer evaluation in the presence of class imbalance. For baseline models, class imbalance is handled via oversampling of minority classes during training. Statistical significance testing has been conducted for all models and will be reported in the results section.

Classical machine learning baselines are optimized using grid search over the parameter ranges listed in Table~\ref{tab:ml_params}, with F1-macro as the scoring metric.
Hyperparameters for all deep learning models, including the number of layers, hidden dimensions, learning rates, and early stopping, are tuned via internal validation within each fold. The search ranges are summarized in Table~\ref{tab:dl_params}. We defined patience as 30 and max epochs as 500.
The final reported results correspond to the mean test F1 scores aggregated over the five external folds, with the standard deviation computed across folds. Detailed results, including statistical significance comparisons between models, are presented in Section~\ref{sec:Results}.

\begin{table}[h!]
\centering
\begin{tabular}{l l l}
\hline
\textbf{Model} & \textbf{Parameter} & \textbf{Values} \\
\hline
Linear & C & [0.01, 0.1, 1, 10] \\
Decision Tree & max depth & [3, 5] \\
              & min samples split & [2, 5, 10] \\
Random Forest & num estimators & [50, 100] \\
              & max depth & [2, 5] \\
              & min samples split & [2, 5] \\
KNN & num neighbors & [3, 5, 7] \\
    & weights & ['uniform', 'distance'] \\
MLP & hidden layer sizes & [(50,), (100,), (100,50)] \\
    & alpha & [0.0001, 0.001] \\
Random Classifier & - & - \\
\hline
\end{tabular}
\caption{Grid search parameters for classical machine learning models.}
\label{tab:ml_params}
\end{table}

\begin{table}[h!]
\centering
\begin{tabular}{l l l}
\hline
\textbf{Model} & \textbf{Parameter} & \textbf{Values} \\
\hline
MLP Deepset & mlp hidden sizes & [75, 150] \\
           & mlp num layers & [2, 3, 4] \\
           & phi num layers & [2, 3] \\
           & rho num layers & [2, 3] \\
           & activation & ['LeakyReLU', 'Relu'] \\
           & lr & [0.005, 0.001] \\
\hline
GNN Deepset & gnn hidden sizes & [75, 150] \\
           & gnn depth & [3, 4] \\
           & phi num layers & [2, 3] \\
           & rho num layers & [2, 3] \\
           & activation & ['LeakyReLU', 'Relu'] \\
           & lr & [0.005, 0.001] \\
\hline
MLP Pooling & mlp hidden sizes & [75, 150] \\
           & mlp num layers & [2, 3, 4] \\
           & activation & ['LeakyReLU', 'Relu'] \\
           & lr & [0.005, 0.001] \\
\hline
GNN Pooling & gnn hidden sizes & [75, 150] \\
           & gnn depth & [1, 2, 3] \\
           & activation & ['LeakyReLU', 'Relu'] \\
           & lr & [0.005, 0.001] \\
\hline
\end{tabular}
\caption{Hyperparameter search ranges for deep learning models.}
\label{tab:dl_params}
\end{table}

\subsection{JamShield Implementation}\label{sec:jamshield_exp}
In this section, we describe the practical implementation of the JamShield defensive framework, detailing both feature-level and data-level interventions, and the choices we made to make the evaluation realistic and reproducible.

\subsubsection{Feature-level defenses: noise and ablation}\label{Feature-leveldefensesImplementation}
For feature-level defenses, we focused on modifying only the subset of playlist features that can be safely perturbed, corresponding to Spotify-calculated attributes: \emph{acousticness}, \emph{danceability}, \emph{instrumentalness}, \emph{liveness}, \emph{loudness}, \emph{popularity art}, \emph{popularity art unique}, \emph{popularity songs}, \emph{ratio unpopulart artists}, \emph{ratio unpopulart artists unique}, \emph{speechiness}, \emph{tempo}, \emph{valence}, and \emph{year add}. 

To introduce noise, we added a Gaussian perturbation with a standard deviation equal to that of the original feature (i.e., \(\pm 3\sigma\)), which represents a substantial variation relative to the natural variability of the data. This choice ensures that the perturbation is significant enough to potentially confuse an attacker while remaining realistic. Each perturbed value is clipped to lie within the original feature's minimum and maximum observed values to prevent unrealistic outliers that could break the plausibility of the playlist. Formally, given a playlist embedding \(\mathbf{p}_i^{(u)}\), the noisy embedding is
\[
\hat{\mathbf{p}}_i^{(u)} \;=\; \mathbf{p}_i^{(u)} + \boldsymbol{\delta}_i^{(u)}, \qquad \|\boldsymbol{\delta}_i^{(u)}\| \le 3\sigma,
\]
with clipping applied component-wise. 

For feature ablation, the subset of features \(\xi^{(u)}\) defined by the Spotify-calculated attributes is removed entirely from the publicly released embeddings. While noise preserves the general structure of the embedding, ablation removes information completely from the selected dimensions.

As we will report in our experiments, feature-level defenses will generally be insufficient to significantly degrade the attacker's performance. Although the Gaussian perturbation of \(3\sigma\) magnitude is large and the ablation removes completely the information, the attacker’s models are resilient to such realistic variations, highlighting the need for stronger data-level interventions.

\subsubsection{Data-level defenses: adversarial playlist injection}\label{sec:Data-level defenses}
Before introducing the two feature-level defense mechanisms, it is essential to clarify which playlist features can be legitimately modified. In our dataset, we identified four categories of features: (i) \emph{Spotify-calculated features} (e.g., acousticness, danceability, tempo, valence), (ii) \emph{Spotify-provided genre labels} (e.g., metal, hip hop), (iii) \emph{public Spotify metadata} (e.g., artist followers, unique artist followers), and (iv) \emph{features reconstructable by the attacker} (e.g., number of artists, playlist duration, release year).
Only the first category---Spotify-calculated audio attributes---can be safely perturbed or removed. Two considerations motivate this restriction. First, the remaining three categories can be easily recovered by an informed adversary through public APIs or external metadata sources, making any modification ineffective. Second, altering genres, public metadata, or reconstructable attributes would not simply reduce information quality, but would instead introduce factual inconsistencies (e.g., wrong release year or incorrect artist statistics), resulting in a clear service disruption. In contrast, small variations or partial removal of Spotify-calculated audio attributes represent a more realistic and acceptable modification within an API design, as they degrade precision without contradicting ground-truth metadata.

The core component of JamShield is the generation of adversarial playlists  (\textit{decoy playlists}), which operates by augmenting the user’s publicly visible set of playlists with carefully selected decoy items. Intuitively, JamShield exploits the fact that certain playlists are statistically associated with specific demographic or behavioural attributes: for example, if a playlist is predominantly listened to by female users, injecting it into the profile of a male user pushes the attacker toward an incorrect prediction. The same reasoning applies to age, personality traits, and other attributes, allowing the defender to strategically steer the model away from the true class.
In practice, we choose these injected playlists from existing items in our dataset rather than generating synthetic ones. This design choice provides a conservative lower-bound estimate of JamShield’s effectiveness: real playlists may be less optimal than tailor-made synthetic decoys, but they guarantee full realism and avoid introducing implausible artifacts.

We frame injection as a constrained optimization directly targeted to the specific attacker model. In our evaluation, \(f^\star\) corresponds to the ten inference pipelines on which our attack achieves the strongest performance (nine GNN–DeepSet variants and one MLP–DeepSet baseline). For a given attacker \(f^\star\) and the true attribute \(T(u)\) of user \(u\), JamShield selects an injection set \(\mathcal{J}(u)\) that maximizes the attacker’s loss under plausibility constraints \(\mathcal{C}\). In our setup, the only constraint we impose is that injected playlists must be drawn from the real dataset, thus ensuring realism by construction:
\begin{equation}\label{eq:injection_obj_targeted}
\mathcal{J}^\star(u) \;=\; 
\arg\max_{\mathcal{J}\,:\,\mathcal{C}(\mathcal{J},u)\le 0} 
L\big(f^\star(\Phi(\mathcal{P}(u)\cup\mathcal{J})), T(u)\big).
\end{equation}

In practice, JamShield operates as follows. For each user, we inject a fixed number \(n\) of playlists, chosen from the real playlists available in the dataset. The defender evaluates each candidate playlist by temporarily adding it to the user's profile and measuring how much it distorts the attacker's prediction. This yields a model-aware ranking of candidates: playlists that cause the largest degradation in the attacker's output are preferred, ensuring that the injected set is inherently confounding for the specific inference model under consideration.

To avoid consistently reusing the same items across users, we apply a lightweight frequency penalty that downweights playlists that have already been selected multiple times. Once a playlist is chosen, it is duplicated as a new node in the user's playlist graph, and all its associated edges are copied so that the relational structure of the graph is updated accordingly.

After the \(n\) playlists have been injected, JamShield performs a white-box refinement stage (since the attacker models are known). Because the attacker model \(f^\star\) is assumed to be known, we apply a bounded projected gradient descent (PGD) update to the duplicated playlists. The refinement is targeted toward the class that is most confounding for the user's true attribute, meaning that even small, realistic perturbations to the duplicated embeddings are explicitly optimized to increase the attacker's loss. This allows JamShield to preserve the plausibility of all injected playlists while effectively altering the graph-level signature consumed by the inference pipeline.

Injection offers several advantages over pure feature-level interventions. By acting at the level of the released playlist set, it directly modifies the input distribution processed by set- and graph-based inference pipelines (i.e., DeepSet and GNN aggregators), while preserving per-playlist coherence so that injected items remain plausible to human observers. This targeted yet realistic procedure allows JamShield to effectively degrade attribute inference performance while remaining deployable on real music-streaming platforms.

%% file: sections/07-Results.tex
\section{Results and discussion}
\label{sec:Results}

This section presents the empirical evaluation of the MusicPIIrate attack and of the proposed countermeasure JamShield. All results follow the experimental protocol described in Section~\ref{sec:ExperimentalSettings}, using macro F1 as the main metric under a user-level 5-fold cross-validation scheme with repeated runs. We report the performance of the three modeling paradigms examined in this work---sample-based, set-based, and graph-based classifiers—and analyze the statistical significance of their differences.

\subsection{MusicPIIrate Results}
\label{subsec:AttackResults}

We first evaluate the ability of MusicPIIrate to infer sensitive user attributes from playlists. The complete results for demographics, habits, and personality traits are reported in Tables~\ref{tab:class-demographics}, \ref{tab:class-habits}, and~\ref{tab:class-personalities}.

\begin{table}[htbp]
\centering
\caption{Mean and std of F1-score in test for model for Demographics.}
\label{tab:class-demographics}
\begin{tabularx}{0.95\linewidth}{l >{\centering\arraybackslash}X >{\centering\arraybackslash}X >{\centering\arraybackslash}X >{\centering\arraybackslash}X >{\centering\arraybackslash}X >{\centering\arraybackslash}X}
\toprule
Model & Age & Country & Econ. & Gender & Marital & Occup. \\
\midrule
Rand. Cl. & 0.06\scriptsize{±.00} & 0.04\scriptsize{±.00} & 0.13\scriptsize{±.00} & 0.27\scriptsize{±.00} & 0.40\scriptsize{±.00} & 0.32\scriptsize{±.00} \\
\midrule
Linear           & 0.36\scriptsize{±.04} & 0.15\scriptsize{±.01} & 0.34\scriptsize{±.03} & 0.49\scriptsize{±.04} & 0.56\scriptsize{±.03} & \textbf{\textcolor{red}{0.65\scriptsize{±.02}}} \\
Dec. Tree     & 0.33\scriptsize{±.06} & 0.08\scriptsize{±.03} & 0.34\scriptsize{±.06} & 0.44\scriptsize{±.04} & 0.55\scriptsize{±.04} & 0.57\scriptsize{±.05} \\
Rand. For.     & 0.35\scriptsize{±.04} & 0.10\scriptsize{±.01} & 0.35\scriptsize{±.05} & 0.46\scriptsize{±.02} & 0.58\scriptsize{±.02} & 0.61\scriptsize{±.04} \\
KNN              & 0.29\scriptsize{±.03} & 0.13\scriptsize{±.02} & 0.33\scriptsize{±.04} & 0.47\scriptsize{±.02} & 0.55\scriptsize{±.03} & 0.58\scriptsize{±.03} \\
MLP              & 0.36\scriptsize{±.03} & 0.15\scriptsize{±.03} & 0.32\scriptsize{±.02} & 0.42\scriptsize{±.03} & 0.57\scriptsize{±.04} & 0.63\scriptsize{±.04} \\
\midrule
MLP Pool.       & 0.35\scriptsize{±.04} & 0.12\scriptsize{±.02} & 0.36\scriptsize{±.03} & 0.45\scriptsize{±.03} & 0.53\scriptsize{±.03} & 0.48\scriptsize{±.03} \\
MLP DS       & 0.38\scriptsize{±.04} & 0.15\scriptsize{±.01} & 0.37\scriptsize{±.02} & 0.46\scriptsize{±.02} & 0.55\scriptsize{±.02} & 0.51\scriptsize{±.02} \\
\midrule
GNN Pool.       & 0.31\scriptsize{±.01} & 0.15\scriptsize{±.01} & 0.35\scriptsize{±.01} & 0.47\scriptsize{±.02} & 0.51\scriptsize{±.01} & 0.51\scriptsize{±.02} \\
GNN DS       & \textbf{\textcolor{red}{0.39\scriptsize{±.02}}} & \textbf{\textcolor{red}{0.18\scriptsize{±.01}}} & \textbf{\textcolor{red}{0.40\scriptsize{±.01}}} & \textbf{\textcolor{red}{0.50\scriptsize{±.02}}} & \textbf{\textcolor{red}{0.60\scriptsize{±.02}}} & 0.58\scriptsize{±.02} \\
\bottomrule
\end{tabularx}
\end{table}

\begin{table}[htbp]
\centering
\caption{Mean and std of F1-score in test for model for Habits.}
\label{tab:class-habits}
\begin{tabularx}{\linewidth}{l >{\centering\arraybackslash}X >{\centering\arraybackslash}X >{\centering\arraybackslash}X >{\centering\arraybackslash}X}
\toprule
Model & Alchol & Smoke & Sport & S.Premium \\
\midrule
Rand. Cl. & 0.31\scriptsize{±.00} & 0.44\scriptsize{±.00} & 0.16\scriptsize{±.00} & 0.20\scriptsize{±.00} \\
\midrule
Linear           & 0.57\scriptsize{±.04} & 0.60\scriptsize{±.03} & 0.31\scriptsize{±.03} & 0.53\scriptsize{±.07} \\
Dec. Tree     & 0.58\scriptsize{±.04} & 0.57\scriptsize{±.03} & 0.26\scriptsize{±.03} & 0.54\scriptsize{±.05} \\
Rand. For.     & 0.54\scriptsize{±.04} & 0.60\scriptsize{±.06} & 0.28\scriptsize{±.04} & \textbf{\textcolor{red}{0.56\scriptsize{±.03}}} \\
KNN              & 0.52\scriptsize{±.02} & \textbf{\textcolor{red}{0.61\scriptsize{±.04}}} & 0.32\scriptsize{±.03} & 0.54\scriptsize{±.03} \\
MLP              & 0.58\scriptsize{±.03} & 0.53\scriptsize{±.05} & 0.34\scriptsize{±.02} & 0.53\scriptsize{±.05} \\
\midrule
MLP Pool.       & 0.53\scriptsize{±.01} & 0.56\scriptsize{±.01} & 0.33\scriptsize{±.01} & 0.40\scriptsize{±.03} \\
MLP DS       & 0.50\scriptsize{±.03} & 0.58\scriptsize{±.03} & \textbf{\textcolor{red}{0.35\scriptsize{±.01}}} & 0.41\scriptsize{±.01} \\
\midrule
GNN Pool.       & 0.54\scriptsize{±.03} & 0.52\scriptsize{±.08} & 0.33\scriptsize{±.02} & 0.44\scriptsize{±.04} \\
GNN DS       & \textbf{\textcolor{red}{0.59\scriptsize{±.03}}} & 0.59\scriptsize{±.01} & 0.34\scriptsize{±.01} & 0.47\scriptsize{±.01} \\
\bottomrule
\end{tabularx}
\end{table}

\begin{table}[htbp]
\centering
\caption{Mean and std of F1-score in test for model for Personality traits.}
\label{tab:class-personalities}
\begin{tabularx}{\linewidth}{l >{\centering\arraybackslash}X >{\centering\arraybackslash}X >{\centering\arraybackslash}X >{\centering\arraybackslash}X >{\centering\arraybackslash}X}
\toprule
Model & Open. & Consc. & Extrav. & Agreeab. & Neurot. \\
\midrule
Rand. Cl. & 0.04\scriptsize{±.00} & 0.11\scriptsize{±.00} & 0.20\scriptsize{±.00} & 0.06\scriptsize{±.00} & 0.12\scriptsize{±.00} \\
\midrule
Linear           & 0.31\scriptsize{±.03} & 0.37\scriptsize{±.03} & 0.35\scriptsize{±.03} & 0.29\scriptsize{±.04} & 0.38\scriptsize{±.01} \\
Dec. Tree     & 0.26\scriptsize{±.04} & 0.38\scriptsize{±.04} & 0.33\scriptsize{±.04} & 0.30\scriptsize{±.04} & 0.35\scriptsize{±.04} \\
Rand. For.     & 0.30\scriptsize{±.03} & 0.37\scriptsize{±.06} & 0.33\scriptsize{±.06} & 0.32\scriptsize{±.03} & 0.37\scriptsize{±.03} \\
KNN              & 0.32\scriptsize{±.02} & 0.35\scriptsize{±.05} & 0.35\scriptsize{±.03} & 0.33\scriptsize{±.03} & 0.34\scriptsize{±.05} \\
MLP              & \textbf{\textcolor{red}{0.35\scriptsize{±.02}}} & 0.35\scriptsize{±.03} & 0.35\scriptsize{±.03} & \textbf{\textcolor{red}{0.35\scriptsize{±.04}}} & 0.39\scriptsize{±.01} \\
\midrule
MLP Pool.       & 0.24\scriptsize{±.05} & 0.38\scriptsize{±.01} & 0.37\scriptsize{±.01} & 0.27\scriptsize{±.04} & 0.39\scriptsize{±.02} \\
MLP DS       & 0.25\scriptsize{±.02} & 0.40\scriptsize{±.01} & 0.39\scriptsize{±.01} & 0.29\scriptsize{±.02} & 0.40\scriptsize{±.00} \\
\midrule
GNN Pool.       & 0.23\scriptsize{±.05} & 0.35\scriptsize{±.02} & 0.34\scriptsize{±.02} & 0.24\scriptsize{±.03} & 0.38\scriptsize{±.01} \\
GNN DS       & 0.25\scriptsize{±.03} & \textbf{\textcolor{red}{0.43\scriptsize{±.01}}} & \textbf{\textcolor{red}{0.42\scriptsize{±.03}}} & 0.28\scriptsize{±.03} & \textbf{\textcolor{red}{0.44\scriptsize{±.01}}} \\
\bottomrule
\end{tabularx}
\end{table}

We begin with the sample-based baselines, which rely on per‑playlist representations. Their performance varies substantially across tasks, generally falling below that of the set‑ and graph‑based models. In demographics, for example, linear and tree‑based models achieve moderate F1 scores, but are consistently outperformed by the more expressive aggregations. Similar trends appear in habits and personality traits, where no baseline achieves top performance in a consistent manner. These results confirm that individual playlists alone provide an incomplete view of a user's musical behaviour, limiting the predictive power of sample‑based approaches.

Set-based models, which aggregate a user's playlists before classification, exhibit differentiated behavior depending on the aggregation strategy. MLP Pooling often lowers the average performance within the group, sometimes performing worse than the baseline methods, indicating that simple pooling can dilute relevant information. In contrast, MLP DeepSet consistently outperforms MLP Pooling in nearly all cases, achieving particularly strong results for specific targets such as sport. These observations indicate that treating a user's playlists as a set can yield improvements, but the choice of aggregation mechanism is crucial: learnable DeepSet aggregation captures user-level information more effectively than naive pooling.

Graph-based models demonstrate the strongest overall performance among all tested families. However, even within this group, a distinction must be made between the choice of aggregation: standard pooling and DeepSet lead to very different outcomes. Using a GNN encoder with standard pooling (GNN-Pooling) usually shows low performance, indicating that the GNN alone is insufficient to exploit the graph structure effectively. In contrast, replacing the pooling aggregation with a learnable DeepSet module (GNN-DeepSet) leads to a marked improvement, producing the model with the best average performance.

The comparison among the four deep architectures (MLP-Pooling, MLP-DeepSet, GNN-Pooling, and GNN-DeepSet) allowed us to conduct an important ablation study. Indeed, we confirmed that the observed gains are not attributable to the encoder or the aggregation individually. GNN-Pooling or MLP-DeepSet alone cannot match the performance of GNN-DeepSet, demonstrating that the strengths of the GNN and the DeepSet aggregation are complementary. This analysis underlines that the exceptional performance of GNN-DeepSet arises from the collaboration between structured playlist modeling and learnable set aggregation, providing a significant improvement over all other configurations and establishing a new state-of-the-art.

\begin{figure*}[!h]
\centering
\includegraphics[width=\linewidth]{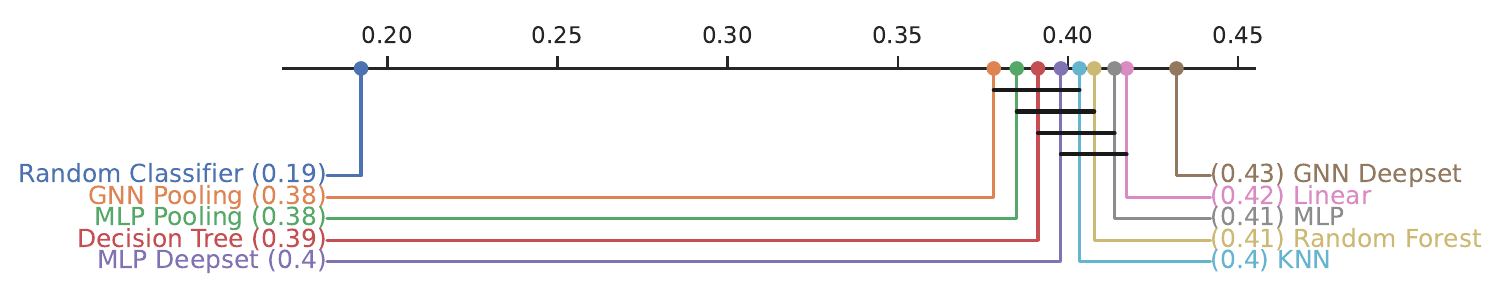}  
\caption{Comparison of CD results for all targets for a p-value of $0.05$. Methods connected by a horizontal bar are not statistically different. As seen, GNN-DeepSet achieves the best average rank and is statistically superior to all competing methods.}
\label{fig:cd_all}
\end{figure*}

To assess whether these differences are statistically significant, we apply a Friedman test followed by Conover post-hoc comparisons with the Holm correction (significance threshold: $p = 0.05$). This approach is particularly suitable in our setting because we have a single dataset with multiple targets: although the prediction label changes, the underlying features remain the same, violating the independence assumption required by other tests. The Friedman test, combined with the Conover post-hoc analysis and Holm correction, provides a robust framework for identifying significant differences while controlling for multiple comparisons. Figure~\ref{fig:cd_all} summarizes the statistical results across all investigated models. Methods connected by a horizontal black bar are not statistically different from each other at the $p = 0.05$ threshold. As shown in the Figure, GNN‑DeepSet achieves the highest performance and its results are statistically superior to all competing methods, confirming that the observed improvement is robust and significant in repeated and targeted runs.


\subsection{JamShield evaluation}

We now evaluate our defenses against the playlist-based inference attack. As explained in Section~\ref{sec:jamshield_exp}, we distinguish between two types of countermeasures: \emph{Feature-level defenses}, which act by modifying or masking specific playlist features, and \emph{Data-level defenses}, which operate by injecting adversarial playlists to obfuscate the user's profile. Since our goal is to assess the effectiveness of these defenses against the method that we proposed, all results are reported with respect to the ten cases in which our models achieve the highest performance—nine based on GNN–DeepSet (age, alchol, conscientiousness, country, economic, extraversion, gender, marital status, and neuroticism) and one on MLP–DeepSet (sport).

\paragraph{\textbf{Feature-level defenses: Noise and Ablation}}
Here, we discuss the results for noise injection and feature ablation. As explained in Section~\ref{Feature-leveldefensesImplementation}, these two countermeasures have been applied only to Spotify-calculated attributes, which constitute the only subset of features safely modifiable/removable without compromising the functionality of the Spotify API. The results show that the performance of our attack remains remarkably stable under both interventions. Across all targets, the reductions in F1 are minimal and never approach a meaningful degradation of the attack.
Although these countermeasures did not produce good performance, this result is still important. Indeed, feature-level perturbations represent the most intuitive mitigation strategy: they are simple to deploy and might be minimally invasive, and thus constitute the natural starting point for designing a privacy defense. Therefore, showing that these naive strategies achieve low results is really important and demonstrates that the countermeasure task is not trivial.
These findings establish a crucial baseline: meaningful protection cannot be achieved by manipulating playlist features alone, thereby further motivating the need for the data-level mechanism introduced next.

\paragraph{\textbf{Data-level defenses: Adversarial Playlist Injection}}
The core of JamShield is based on adversarial playlist injection, whereby additional playlists are added to the user profile to make the attacker model missclassify the right class of a user. Table~\ref{tab:plnoise_summary} reports the mean F1 scores achieved by our attack models under different numbers of injected playlists ($K$). Even adding a single playlist per user ($K=1$) produces a noticeable reduction in performance, with an average drop of 10\% across targets. Higher values of $K$ usually further degrade the performance, although for others the decline quickly stabilises.
We limited the number of injected decoy playlists to four, as larger values would reduce the practical realism of the defense. Although injecting more playlists could, in principle, yield a stronger protective effect, doing so would risk degrading the experience for API consumers, thus making high values of $K$ impractical to deploy in real systems.

It is also worth noting that the results are not expected to drop to the level of a random classifier. As discussed in the previous section, our injected playlists are real items taken from the dataset, not synthetic or adversarially optimised playlists specifically engineered to break the attack. This design choice makes our evaluation more realistic: the reductions we observe may not represent the maximum theoretical effect achievable from JamShield, but they guarantee a good protection for our users, being at the same time statistically significant, robust, and conservative.

\begin{table*}[!h]
\centering
\resizebox{\textwidth}{!}{%
\begin{tabular}{lcccccccccc}
\toprule
& age & alchol & conscientiousness & country & economic & extraversion & gender & marital\_status & neuroticism & sport \\
\midrule
MusicPIIrate
& \( 0.38 \pm 0.02 \) & \( 0.59 \pm 0.03 \) & \( 0.43 \pm 0.00 \) & \( 0.18 \pm 0.01 \) & \( 0.38 \pm 0.01 \) & \( 0.41 \pm 0.02 \) & \( 0.50 \pm 0.02 \) & \( 0.60 \pm 0.02 \) & \( 0.43 \pm 0.01 \) & \( 0.34 \pm 0.02 \) \\
\midrule

Rand. Cl.
& \( 0.06 \pm 0.00 \) & \( 0.31 \pm 0.00 \) & \( 0.11 \pm 0.00 \) & \( 0.04 \pm 0.00 \) & \( 0.13 \pm 0.00 \) & \( 0.20 \pm 0.00 \) & \( 0.27 \pm 0.00 \) & \( 0.40 \pm 0.00 \) & \( 0.12 \pm 0.00 \) & \( 0.16 \pm 0.00 \) \\
(avg -0.29) & (-0.32) & (-0.28) & (-0.32) & (-0.14) & (-0.25) & (-0.21) & (-0.23) & (-0.20) & (-0.31) & (-0.18) \\
\midrule

JamShield 1
& \( 0.26 \pm 0.03 \) & \( 0.44 \pm 0.06 \) & \( 0.33 \pm 0.01 \) & \( 0.13 \pm 0.01 \) & \( 0.29 \pm 0.01 \) & \( 0.31 \pm 0.02 \) & \( 0.37 \pm 0.01 \) & \( 0.47 \pm 0.05 \) & \( 0.29 \pm 0.03 \) & \( 0.31 \pm 0.01 \) \\
(avg -0.10) & (-0.12) & (-0.15) & (-0.10) & (-0.05) & (-0.09) & (-0.10) & (-0.13) & (-0.13) & (-0.14) & (-0.03) \\

JamShield 2 \rule{0pt}{2.5ex}
& \( 0.22 \pm 0.03 \) & \( 0.40 \pm 0.03 \) & \( 0.32 \pm 0.02 \) & \( 0.11 \pm 0.01 \) & \( 0.29 \pm 0.02 \) & \( 0.32 \pm 0.02 \) & \( 0.37 \pm 0.03 \) & \( 0.47 \pm 0.04 \) & \( 0.30 \pm 0.01 \) & \( 0.30 \pm 0.01 \) \\
(avg -0.11) & (-0.16) & (-0.19) & (-0.11) & (-0.07) & (-0.09) & (-0.09) & (-0.13) & (-0.13) & (-0.13) & (-0.04) \\

JamShield 4 \rule{0pt}{2.5ex}
& \( 0.19 \pm 0.05 \) & \( 0.43 \pm 0.05 \) & \( 0.32 \pm 0.02 \) & \( 0.11 \pm 0.02 \) & \( 0.29 \pm 0.01 \) & \( 0.27 \pm 0.05 \) & \( 0.36 \pm 0.02 \) & \( 0.49 \pm 0.04 \) & \( 0.29 \pm 0.01 \) & \( 0.30 \pm 0.02 \) \\
(avg -0.12) & (-0.19) & (-0.16) & (-0.11) & (-0.07) & (-0.09) & (-0.14) & (-0.14) & (-0.11) & (-0.14) & (-0.04) \\
\bottomrule
\end{tabular}
}
\caption{The table shows the average results for JamShield. The first two rows report the F1-score for MusicPIIrate (the baseline) and for the random classifier. The following rows report the F1-score and the mean difference for task and for number of added playlist (i.e., 1, 2, and 4 playlists added).}

\label{tab:plnoise_summary}
\end{table*}

\subsection{Limitations}
\label{subsec:Limitations}
Our primary limitation lies in the scope of our data, as the MusicPIIrate framework was exclusively evaluated using a dataset derived from the Spotify platform. This constraint is not unique to our work but is a recognized difficulty in state-of-the-art (SOTA) web analyses, where obtaining diverse data is inherently challenging due to stringent platform restrictions and access policies~\cite{pajola2025elephant}. Consequently, while our findings demonstrate a significant privacy vulnerability within the tested environment, future work is necessary to validate the generalizability of these PII inference capabilities across other music streaming services.

Despite the strong performance of GNN-DeepSet, no single model consistently dominates habits or personality traits. This variability suggests that some tasks might be sufficiently simple or low-dimensional that simpler models already might capture most of the available information.
Last, the playlist-level countermeasure. With a larger dataset, JamShield could have potentially selected a more diverse set of decoy playlists, possibly leading to a stronger degradation of the attacker’s performance. Therefore, the current results provide a conservative estimate of the countermeasure’s effectiveness, and larger datasets may enable even more robust protection.

Another limitation of our framework is its assumption of a static dataset, in which user playlists are treated as temporally invariant. In real-world scenarios user behavior is inherently dynamic and may evolve, leading to potential behavioral drift that could degrade the model’s predictive performance. This limitation could be mitigated through periodic retraining or online adaptation strategies. A promising direction for future work could be the exploration of continuous learning techniques to manage ever-changing user profiles.

%% file: sections/08-Conclusions.tex
\subsection{Conclusions}

This work demonstrates that publicly shared music playlists, which are perceived as harmless expressions of taste, encode sufficient information to enable accurate and systematic inference of sensitive personal attributes.
To investigate this threat, we propose MusicPIIrate, a novel offensive framework that leverages cutting-edge deep learning techniques to jointly model playlist-level relationships and the set-based structure of user collections.
Our evaluation shows that combining Graph Neural Networks with DeepSets consistently outperforms the baseline models, as confirmed by the statistical test with a significance level of $p=0.05$. Thanks to these findings, GNN–DeepSet establishes a novel state-of-the-art standard for PII inference from playlist music data.

To counter this threat, we proposed JamShield, a practical and realistically deployable defensive framework. Our results highlight that simple feature-level perturbations provide negligible protection, whereas data-level adversarial playlist injection substantially reduces the attacker’s performance. Indeed, even a single injected playlist lowers F1 by 10\% on average, while four playlists further degrade the attack, effectively mitigating privacy risks without compromising the usability of the modified API.

Overall, this work reveals both the concreteness of the privacy risks associated with public playlists and the feasibility of deploying meaningful countermeasures. By providing a rigorous offensive benchmark and a validated defence, we aim to inform the design of privacy-aware music streaming platforms and support future research on safeguarding users against inference-based attacks in data-rich digital ecosystems.

%% file: bibliography.bib
@inproceedings{schroer2025sok,
  title={SoK: On the offensive potential of AI},
  author={Schr{\"o}er, Saskia Laura and Apruzzese, Giovanni and Human, Soheil and Laskov, Pavel and Anderson, Hyrum S and Bernroider, Edward WN and Fass, Aurore and Nassi, Ben and Rimmer, Vera and Roli, Fabio and others},
  booktitle={2025 IEEE Conference on Secure and Trustworthy Machine Learning (SaTML)},
  pages={247--280},
  year={2025},
  organization={IEEE}
}

@online{IC3_2024,
  author={{Internet Crime Complaint Center (IC3)}},
  title={Criminals Use Generative Artificial Intelligence to Facilitate Financial Fraud},
  year={2024},
  url={https://www.ic3.gov/PSA/2024/PSA241203},
  note={Alert Number: I-120324-PSA, December 3, 2024}
}

@inproceedings{tricomi2024all,
  title={`All of Me': Mining Users' Attributes from their Public Spotify Playlists},
  author={Tricomi, Pier Paolo and Pajola, Luca and Pasa, Luca and Conti, Mauro},
  booktitle={Companion Proceedings of the ACM Web Conference 2024},
  pages={963--966},
  year={2024}
}

@inproceedings{hu2022generating,
  title={Generating adversarial malware examples for black-box attacks based on GAN},
  author={Hu, Weiwei and Tan, Ying},
  booktitle={International Conference on Data Mining and Big Data},
  pages={409--423},
  year={2022},
  organization={Springer}
}

@inproceedings{adali2012predicting,
  title={Predicting personality with social behavior},
  author={Adali, Sibel and Golbeck, Jennifer},
  booktitle={2012 IEEE/ACM International Conference on Advances in Social Networks Analysis and Mining},
  pages={302--309},
  year={2012},
  organization={IEEE}
}

@article{gong2018attribute,
  title={Attribute inference attacks in online social networks},
  author={Gong, Neil Zhenqiang and Liu, Bin},
  journal={ACM Transactions on Privacy and Security (TOPS)},
  volume={21},
  number={1},
  pages={1--30},
  year={2018},
  publisher={ACM New York, NY, USA}
}

@article{farid2022creating,
  title={Creating, using, misusing, and detecting deep fakes},
  author={Farid, Hany},
  journal={Journal of Online Trust and Safety},
  volume={1},
  number={4},
  year={2022}
}

@inproceedings{pajola2025phishgen,
title={E-PhishGen: Unlocking Novel Research in Phishing Email Detection},
author={Pajola, Luca and Caripoti, Eugenio and Pizzi, Simeone and Conti, Mauro and Banzer, Stefan and Apruzzese, Giovanni},
booktitle={ACM Workshop on Artificial Intelligence Security (AISec)},
year={2025}
}

@inproceedings{pajola2025elephant,
  title={Elephant in the Room: Dissecting and Reflecting on the Evolution of Online Social Network Research},
  author={Pajola, Luca and Schr{\"o}er, Saskia Laura and Tricomi, Pier Paolo and Conti, Mauro and Apruzzese, Giovanni},
  booktitle={Proceedings of the International AAAI Conference on Web and Social Media},
  volume={19},
  pages={1436--1452},
  year={2025}
}

@article{schroer2025exploiting,
  title={Exploiting AI for Attacks: On the Interplay between Adversarial AI and Offensive AI},
  author={Schr{\"o}er, Saskia Laura and Pajola, Luca and Castagnaro, Alberto and Apruzzese, Giovanni and Conti, Mauro},
  journal={IEEE Intelligent Systems},
  year={2025}
}

@inproceedings{Zaheer2017,
author = {Zaheer, Manzil and Kottur, Satwik and Ravanbhakhsh, Siamak and P{\'{o}}czos, Barnab{\'{a}}s and Salakhutdinov, Ruslan and Smola, Alexander J.},
booktitle = {Advances in Neural Information Processing Systems},
pages = {3391--3401},
title = {{Deep Sets}},
year = {2017}
}

@article{micheli2009neural,
  title={Neural network for graphs: A contextual constructive approach},
  author={Micheli, Alessio},
  journal={IEEE Transactions on Neural Networks},
  volume={20},
  number={3},
  pages={498--511},
  year={2009},
  publisher={IEEE}
}

@inproceedings{kipf2016semi,
archivePrefix = {arXiv},
arxivId = {1609.02907},
author = {Kipf, Thomas N. and Welling, Max},
booktitle = {ICLR},
doi = {10.1051/0004-6361/201527329},
eprint = {1609.02907},
isbn = {9781611970685},
issn = {0004-6361},
pages = {1--14},
title = {{Semi-Supervised Classification with Graph Convolutional Networks}},
url = {http://arxiv.org/abs/1609.02907},
year = {2017}
}

@inproceedings{Navarin2019,
address = {Budapest, Hungary},
author = {Navarin, Nicol{\`{o}} and Tran, Dinh Van and Sperduti, Alessandro},
booktitle = {International Joint Conference on Neural Networks},
title = {{Universal Readout for Graph Convolutional Neural Networks}},
year = {2019}
}

@inproceedings{weinsberg2012blurme,
  title={BlurMe: Inferring and obfuscating user gender based on ratings},
  author={Weinsberg, Udi and Bhagat, Smriti and Ioannidis, Stratis and Taft, Nina},
  booktitle={Proceedings of the sixth ACM conference on Recommender systems},
  pages={195--202},
  year={2012}
}

@inproceedings{tricomi2023attribute,
  title={Attribute inference attacks in online multiplayer video games: A case study on Dota2},
  author={Tricomi, Pier Paolo and Facciolo, Lisa and Apruzzese, Giovanni and Conti, Mauro},
  booktitle={Proceedings of the Thirteenth ACM Conference on Data and Application Security and Privacy},
  pages={27--38},
  year={2023}
}

@article{rentfrow2003re,
  title={The do re mi's of everyday life: the structure and personality correlates of music preferences.},
  author={Rentfrow, Peter J and Gosling, Samuel D},
  journal={Journal of personality and social psychology},
  volume={84},
  number={6},
  pages={1236},
  year={2003},
  publisher={American Psychological Association}
}

@article{rentfrow2012role,
  title={The role of music in everyday life: Current directions in the social psychology of music},
  author={Rentfrow, Peter J},
  journal={Social and personality psychology compass},
  volume={6},
  number={5},
  pages={402--416},
  year={2012},
  publisher={Wiley Online Library}
}

@article{chamorro2007personality,
  title={Personality and music: Can traits explain how people use music in everyday life?},
  author={Chamorro-Premuzic, Tomas and Furnham, Adrian},
  journal={British journal of psychology},
  volume={98},
  number={2},
  pages={175--185},
  year={2007},
  publisher={Wiley Online Library}
}

@book{north2008social,
  title={The social and applied psychology of music},
  author={North, Adrian and Hargreaves, David},
  year={2008},
  publisher={OUP Oxford}
}

@article{sloboda2001emotions,
  title={Emotions in everyday listening to music},
  author={Sloboda, John A and O’neill, Susan A},
  journal={Music and emotion: Theory and research},
  volume={8},
  pages={415--429},
  year={2001}
}

@inproceedings{liu2012inferring,
  title={Inferring personal traits from music listening history},
  author={Liu, Jen-Yu and Yang, Yi-Hsuan},
  booktitle={Proceedings of the second international ACM workshop on Music information retrieval with user-centered and multimodal strategies},
  pages={31--36},
  year={2012}
}

@article{krismayer2019predicting,
  title={Predicting user demographics from music listening information},
  author={Krismayer, Thomas and Schedl, Markus and Knees, Peter and Rabiser, Rick},
  journal={Multimedia Tools and Applications},
  volume={78},
  number={3},
  pages={2897--2920},
  year={2019},
  publisher={Springer}
}

@article{anderson2021just,
  title={“Just the way you are”: Linking music listening on Spotify and personality},
  author={Anderson, Ian and Gil, Santiago and Gibson, Clay and Wolf, Scott and Shapiro, Will and Semerci, Oguz and Greenberg, David M},
  journal={Social Psychological and Personality Science},
  volume={12},
  number={4},
  pages={561--572},
  year={2021},
  publisher={Sage Publications Sage CA: Los Angeles, CA}
}

@incollection{golbeck2011predicting,
  title={Predicting personality with social media},
  author={Golbeck, Jennifer and Robles, Cristina and Turner, Karen},
  booktitle={CHI'11 extended abstracts on human factors in computing systems},
  pages={253--262},
  year={2011}
}

@article{kosinski2013private,
  title={Private traits and attributes are predictable from digital records of human behavior},
  author={Kosinski, Michal and Stillwell, David and Graepel, Thore},
  journal={Proceedings of the national academy of sciences},
  volume={110},
  number={15},
  pages={5802--5805},
  year={2013},
  publisher={National Acad Sciences}
}

@article{sust2023personality,
  title={Personality computing with naturalistic music listening behavior: Comparing audio and lyrics preferences},
  author={Sust, Larissa and Stachl, Clemens and Kudchadker, Gayatri and B{\"u}hner, Markus and Schoedel, Ramona},
  journal={Collabra: Psychology},
  volume={9},
  number={1},
  pages={75214},
  year={2023},
  publisher={University of California Press}
}

@inproceedings{sah2025perfairx,
  title={PerFairX: Is There a Balance Between Fairness and Personality in Large Language Model Recommendations?},
  author={Sah, Chandan Kumar and Lian, Xiaoli},
  booktitle={Proceedings of the IEEE/CVF International Conference on Computer Vision},
  pages={2750--2759},
  year={2025}
}

@inproceedings{Keriven_2022,
	author = {Keriven, N.},
	booktitle = {The First Learning on Graphs Conference},
	title = {Not too little, not too much: a theoretical analysis of graph (over)smoothing},
	year = {2022}}

@article{pasa2022,
	author = {Pasa, L. and Navarin, N. and Sperduti, A.},
	journal = {Machine Learning},
	number = {4},
	pages = {1205--1237},
	title = {Polynomial-based graph convolutional neural networks for graph classification},
	volume = {111},
	year = {2022}}

@inproceedings{Wu2019,
	author = {Wu, F. and Zhang, T. and de Souza, A. H. and Fifty, C. and Yu, T. and Weinberger, K. Q.},
	booktitle = {International conference on machine learning},
	title = {Simplifying graph convolutional networks},
	year = {2019}}
